\documentclass{article}
\usepackage{spconf,graphicx}
\usepackage[utf8]{inputenc}
\usepackage[english]{babel}
\usepackage[unicode=true]{hyperref}
\usepackage[T1]{fontenc}
\usepackage{amsmath}

\usepackage[
    subtle, 
    leading,
    lists,
    bibnotes
    ]{savetrees}

\usepackage[sorting=none, giveninits=true, maxbibnames=99, style=numeric-comp]{biblatex}
\bibliography{zoterobib.bib}
\AtEveryBibitem{
    \clearlist{language} 
    \clearfield{urlyear} 
    \clearfield{note} 
    \clearfield{url}
    \clearfield{doi}
}    
\usepackage{csquotes}

\usepackage{placeins}

\usepackage{booktabs}
\usepackage{subcaption}

\title{Multi-Microphone Noise Data Augmentation for DNN-based Own Voice\\Reconstruction for Hearables in Noisy Environments}
\author{}

\name{
Mattes Ohlenbusch$^{\star}$
\thanks{
The Oldenburg Branch for Hearing, Speech and Audio Technology HSA is funded in the program \frqq Vorab\flqq~by the Lower Saxony Ministry of Science and Culture (MWK) and the Volkswagen Foundation for its further development. 
This work was partly funded by the German Ministry of Science and Education BMBF FK 16SV8811 and the Deutsche Forschungsgemeinschaft (DFG, German Research Foundation) - Project ID 352015383 – SFB 1330 C1.
}, 
Christian Rollwage$^\star$, 
Simon Doclo$^{\star,\dagger}$
}

\address{
\small $^{\star}$ Fraunhofer Institute for Digital Media Technology IDMT, Oldenburg Branch for Hearing, Speech and Audio Technology HSA, Germany \\
\small $^{\dagger}$ 
Dept. of Medical Physics and Acoustics and Cluster of Excellence Hearing4all, Carl von Ossietzky Universität Oldenburg, Germany
}

\begin{document}
\maketitle

\begin{abstract} 
Hearables with integrated microphones may offer communication benefits in noisy working environments, e.g. by transmitting the recorded own voice of the user. 
Systems aiming at reconstructing the clean and full-bandwidth own voice from noisy microphone recordings are often based on supervised learning.
Recording a sufficient amount of noise required for training such a system is costly since noise transmission between outer and inner microphones varies individually. Previously proposed methods either do not consider noise, only consider noise at outer microphones or assume inner and outer microphone noise to be independent during training, and it is not yet clear whether individualized noise can benefit the training of and own voice reconstruction system. 
In this paper, we investigate several noise data augmentation techniques based on measured transfer functions to simulate multi-microphone noise.
Using augmented noise, we train a multi-channel own voice reconstruction system.  
Experiments using real noise are carried out to investigate the generalization capability. 
Results show that incorporating augmented noise yields large benefits, in particular considering individualized noise augmentation leads to higher performance. 
\end{abstract} 


\section{Introduction}
In noisy working environments such as industrial production or surgery rooms, communication is often impaired. 
Radio communication based on own voice pickup and transmission can considerably improve communication~\cite{nordholm_assistive_2015}.
In applications where the safety equipment such as breathing masks, protective helmets, or earmuffs prevents using a close-talk microphone in front of the talkers mouth, in-the-ear hearables with integrated microphones are more suited to record the own voice of the talker. 
Both an outer microphone
(OM) and an inner microphone (IM) can be beneficial for systems aiming at reconstructing the own voice of the hearable user from noisy recordings.
However, OM recordings suffer from external noise, and IM recordings suffer from low-frequency amplification and band-limitation (occlusion), and external and body-produced noise.
Therefore, an own voice reconstruction system estimating clean broadband speech from noisy hearable microphone recordings is required to enable communication.
Current approaches usually rely on deep learning, where large amounts of training data are required. 
In addition to relying on the availability of device-specific own voice recordings, external noise has to be accounted for during training. 
For inward facing body-conduction sensors, it is often assumed no external noise is picked up~\cite{wang_fusing_2022, hauret_eben_2023, liu_bone-conducted_2018, yu_time-domain_2020}. 
While other multi-microphone DNN-based systems for hearing device speech enhancement are often trained using artificial head impulse responses for noise, e.g.~\cite{tammen_deep_2022,westhausen_low_2023}, it has previously not been investigated whether this is sufficient for systems utilizing IMs.

In~\cite{hauret_eben_2023, liu_bone-conducted_2018} it has been proposed to reconstruct the own voice from body-conduction sensor recordings using a deep neural network (DNN) without accounting for external noise. 
In~\cite{ohlenbusch_training_2022}, a bandwidth-extension based approach for hearable IM recordings has been trained similarly, but with added body-produced noise recordings during training. 
In~\cite{wang_fusing_2022, yu_time-domain_2020}, multi-channel systems utilizing both an OM and an inward facing body-conduction sensor are proposed where during training, noise is only added to the OM signal. 
In~\cite{tammen_dictionary-based_2022}, a dictionary-based approach has been proposed in which OM and noise at a contact microphone are modeled independently.
In~\cite{wang_multi-modal_2022}, a multi-channel system for automatic own voice speech recognition has been proposed which was trained by adding noise only to the OM signals. 
In~\cite{li_enabling_2023}, real own voice recordings with a bone conduction sensor have been used for training a reconstruction system without accounting for external noise. Fine-tuning with noise recorded from different talkers was investigated. 
For IMs however, previous research suggests that the transmission of external noise through the device does not only depend on the device used, but also on individual differences~\cite{denk_hearpiece_2021} and the direction of arrival~\cite{liebich_direction--arrival_2018}.

In this paper, we propose several techniques to simulate external noise at hearable microphones for use in data augmentation. 
To our knowledge, this is the first work to address noise data augmentation in the context of own voice reconstruction with an OM and an IM. 
The influence of noise augmentation techniques is evaluated based on real speech and noise recordings. 
Results show that the use of the proposed techniques, in particular the use of individualized noise augmentation, leads to superior performance. 
In an ablation study, we find that in low signal-to-noise ratios (SNRs) the contribution of the IM is larger, while in high SNRs high performance can be achieved using only the OM.

\section{Signal Model}
\begin{figure}[t]
    \centering
    \includegraphics[page=3,width=\columnwidth]{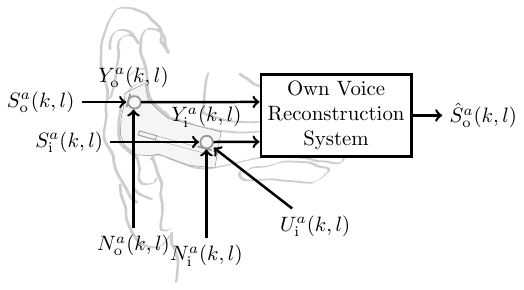}
    \caption{Noisy multi-microphone own voice reconstruction.} 
    \label{fig:sigmodel}
\end{figure}

Fig.~\ref{fig:sigmodel} depicts the considered scenario, where a talker is wearing an in-the-ear hearable device equipped with an OM and an IM, denoted by subscript $\mathrm{o}$ and $\mathrm{i}$, respectively. 
In the short time Fourier transform (STFT) domain, the noisy own voice signal of talker $a$ recorded at the OM is denoted by $Y_\mathrm{o}^a(k,l)$ where $k$ is the frequency bin index and $l$ is the time frame index. 
The recorded noisy own voice signal at the OM consists of an own voice component $S_\mathrm{o}^a(k,l)$ and an external noise component $N_\mathrm{o}^a(k,l)$, i.e.
\begin{equation}
    Y_\mathrm{o}^a(k,l) = S_\mathrm{o}^a(k,l) + N_\mathrm{o}^a(k,l).
\end{equation}
Since the IM is located inside the occluded ear canal, body-produced sounds such as breathing or heartbeats are also recorded~\cite{bouserhal_-ear_2019}. 
Hence, we define the noisy own voice signal recorded at the IM as
\begin{equation}
    Y_\mathrm{i}^a(k,l) = S_\mathrm{i}^a(k,l) + N_\mathrm{i}^a(k,l) + U_\mathrm{i}^a(k,l),
\end{equation}
where $U_\mathrm{i}^a(k,l)$ denotes body-produced noise.
We further assume that the external noise components at the OM and the IM are related by a linear, time-invariant, direction-dependent relative transfer function (RTF) $G_\mathrm{o,i}(k,\theta)$, so that for a single spatially stationary noise source from direction $\theta$ the external IM noise component is 
\begin{equation}
    N_\mathrm{i}^a(k,l) =  N_\mathrm{o}^a(k,l) \cdot G_\mathrm{o,i}^a(k,\theta).
\end{equation} 
For noise fields consisting of several sources, the noise components of each source are assumed to add up to the recorded noise field.

In this work, the goal is to obtain an own voice reconstruction system able to estimate $S_\mathrm{o}^a(k,l)$ from the noisy recordings $Y_\mathrm{o}^a(k,l)$ and $Y_\mathrm{i}^a(k,l)$ obtained from a single hearable device. 
For training such a system, different methods of simulating external noise transmission during training are investigated.
\FloatBarrier

\section{Noise Data Augmentation} 
\label{sec:noise}
Transfer function measurements can be utilized to simulate a large corpus of multi-channel hearable recordings of external noise based on a large single-channel noise corpus. 
Using a single-channel recording at a reference microphone with STFT $N^\mathrm{ref}(k,l)$, the recorded noise at both considered hearable microphones can be modeled using an RTF.
For the OM of a hearable worn by talker $a$, we assume the augmented external noise $\hat{N}_\mathrm{o}^a$ is equal to the single-channel recording, so that
\begin{equation}
   \hat{N}_\mathrm{o}^a(k,l) = N^\mathrm{ref}(k,l).
\end{equation}
For the IM, the augmented noise $\hat{N}_\mathrm{i}^a$ is modeled by the RTF $\hat{G}_\mathrm{o,i}$:

\begin{itemize}
\item \textbf{No IM noise}
Assuming no external noise arrives at the IM, i.e. $\hat{G}_\mathrm{o,i}=0$, the noise component is equal to
\begin{equation}
    \hat{N}_\mathrm{i}^a(k,l) = 0.
    \label{eq:no_noise}
\end{equation}
\item \textbf{Artificial head}
Assuming the RTF depends on the direction of arrival, but neglecting individual differences, we propose to use a set of measurements for different directions $\theta$ using an artificial head (AH):
\begin{equation}
    \hat{N}_\mathrm{i}^a(k,l,\theta) =  \hat{N}_\mathrm{o}^\mathrm{a}(k,l) \cdot \hat{G}_\mathrm{o,i}^\mathrm{AH}(k, \theta).
\end{equation}

\item \textbf{Non-individual}
Accounting only for directional differences but instead of an AH using RTFs from a single talker $b$, we propose to model the noise component as
\begin{equation}
    \hat{N}_\mathrm{i}^a(k,l) = \hat{N}_\mathrm{o}^\mathrm{a}(k,l) \cdot \hat{G}_\mathrm{o,i}^b(k,\theta). 
\end{equation}
\item \textbf{Individual}
Assuming the RTF is subject to individual differences and direction of arrival, both direction-dependent and individual variations can be accounted for by using directional RTFs for each individual talker:  
\begin{equation}
    \hat{N}_\mathrm{i}^a(k,l,\theta) = \hat{N}_\mathrm{o}^\mathrm{a}(k,l) \cdot \hat{G}_\mathrm{o,i}^a(k, \theta).
    \label{eq:indivnoise}
\end{equation}
During speech and noise mixing, $a$ is the same talker an own voice and augmented noise recording to be mixed.
\end{itemize}
In order to simulate single sources for obtaining multi-channel noise data for DNN training, the direction $\theta$ is chosen at random.
Since realistic noisy environments often consist of more than one noise source, we also simulate pseudo-diffuse noise with the direction-dependent methods.
In order to simulate pseudo-diffuse noise, noise signals from all possible source directions $\theta$ are obtained using one of the methods in \eqref{eq:no_noise}-\eqref{eq:indivnoise} and added:
\begin{equation}
    \hat{N}_\mathrm{i}^{a,\mathrm{diff}}(k,l) = \sum_\theta \hat{N}_\mathrm{i}^a(k,l,\theta).
\end{equation}
Each directional reference noise signal is further delayed by one second, so that no two directional signals can be synchronous. 

For both pseudo-diffuse or single-source noise, white noise is added to RTF-based IM noise with a random level in $[-\infty, -60]$\,dB relative to the IM external noise component.
This procedure reduces the coherence between the IM and OM noise signals, bringing it closer to measured coherence of real external noise recordings. 
For each recording it is randomly decided whether a single source noise or pseudo-diffuse noise is obtained with a a probability of $0.5$ each during training with each augmentation method.

\section{Experimental setup}

\subsection{Datasets}
For data augmentation, approximately 180\,h of single-channel noise recordings from the fifth DNS challenge~\cite{dubey_deep_2023} are used.
The AH RTFs are obtained from the Hearpiece database~\cite{denk_hearpiece_2021} where the closed-vent variant of a prototype hearable~\cite{denk_one-size-fits-all_2019} is used. 
Directional RTFs are chosen either as 8 horizontal directions in 45\,°-steps (Artificial head), or with fine resolution in 7.5\,°-steps (Artificial head fine). 
The concha mic. is chosen as the OM. 
From 18 individual talkers wearing the same hearable, external noise RTFs are measured for 8 horizontal directions in 45\,°-steps using exponential sweeps  from 80\,Hz to 22.05\,kHz with a duration of 3\,s played from 8 loudspeakers in a circle with a distance of 1.5\,m around the talker. 
Both sets of RTFs as well as the single-channel noise recordings are only used for simulating training data. 
For non-individual methods, a random talker is chosen, and for the non-individual non-directional method, a random direction is chosen as well.
From the same talkers, individual multi-channel external noise recordings were obtained in the same loudspeaker configuration. 
The following noise types were recorded: single-source surgery room noise, metal grinder, directional babble, pseudo-diffuse babble, pseudo-diffuse surgery room noise, pseudo-diffuse factory noise. 
These real noise recordings are only used for testing.
From the same talkers, approximately 25-30 minutes of German own voice speech per talker were recorded in a sound-proofed listening booth while they were wearing the hearable devices. 
As these recordings are obtained in-situ, the body-produced noises are also recorded at the IM. 
Recordings are split into train/validation/test parts consisting of 12/2/4 talkers without overlap.

\subsection{Training details}
Audio files are resampled to $16$\,kHz.
Own voice utterances are cut to 3\,s length.
In this work, only recordings from the left ear device are used.
Speech and noise recordings are mixed to a range of [-10,~25]\,dB SNR defined at the OM, and the IM noise is scaled accordingly so that noise level differences are preserved.
Mean-variance normalization is applied to the noisy signal for each microphone individually.
The noisy signal statistics of the OM are also utilized to scale the target speech signal by the same amount as the speech component in the noisy signal as in~\cite{braun_data_2020}.
STFTs are computed with frame size of 512 samples corresponding to 32\,ms and 50\,\% overlap, where both in analysis and synthesis a square-root Hann window is used.
A batch size of 4 is used in training. 
We utilize the combined $L_1$ loss of the time-domain and STFT-domain estimated and target speech signals~\cite{wang_stft-domain_2023}.
The Adam optimizer~\cite{kingma_adam_2015} with learning rate $10^{-4}$ is used.
Training is carried out to a maximum of 100 epochs. 
The learning rate is halved after 3 consecutive epochs without validation loss improvement and early stopping is applied after 6 consecutive epochs without improvement. 
For the noise augmentation experiment, the OM+IM DNN variant is trained with real own voice recordings and external noise obtained from using the different augmentation methods described in Section~\ref{sec:noise}. 
Testing is carried out using real own voice recordings and real external noise recordings.
Trained system performance is evaluated in terms of wideband PESQ~\cite{international_telecommunications_union_itu_itu-t_2001} and STOI~\cite{taal_algorithm_2011}. 
The clean OM speech signal is used as reference for both metrics. 
Results are averaged over talkers and noise types.

\begin{figure}[t]
    \centering
    \includegraphics[width=\columnwidth,page=1]{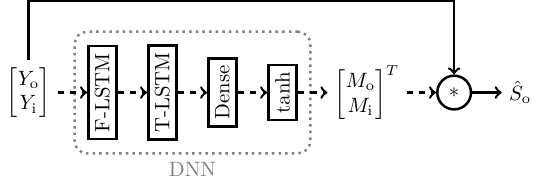}
    \caption{DNN-based own voice reconstruction system utilizing outer and inner microphone for mask estimation and filtering (OM+IM) based on the FT-JNF architecture from~\cite{tesch_insights_2023}.}
    \label{fig:dnn_architecture}
\end{figure}

\begin{table}[t]
    \centering
    \begin{tabular*}{\columnwidth}{|@{\extracolsep{\fill} } l|l|l|l|}
    \hline 
       \textbf{DNN} & \textbf{Input} & \textbf{Output} & \textbf{Own voice estimate} \\ \hline
       OM   & $Y_\mathrm{o}$ &  $M_\mathrm{o}$ & $\hat{S}_\mathrm{o} = M_\mathrm{o} \cdot Y_\mathrm{o}$\\
       IM   & $Y_\mathrm{i}$ &  $M_\mathrm{i}$ & $\hat{S}_\mathrm{o} = M_\mathrm{i} \cdot Y_\mathrm{i}$ \\
       OM+auxIM & $Y_\mathrm{o}, Y_\mathrm{i}$ & $M_\mathrm{o}$ & $\hat{S}_\mathrm{o} = M_\mathrm{o} \cdot Y_\mathrm{o}$ \\
       OM+IM & $Y_\mathrm{o}, Y_\mathrm{i}$ & $M_\mathrm{o}, M_\mathrm{i}$ & $\hat{S}_\mathrm{o} = \sum\limits_{m\in\{\mathrm{i,o}\}} M_\mathrm{m} \cdot Y_\mathrm{m}$ \\
       \hline 
    \end{tabular*}
    \caption{
    DNN variants with different microphone contributions to mask estimation and STFT filtering.  
    Here, auxIM indicates the IM is only used as auxiliary input for mask estimation, but not as a signal to be filtered. 
    STFT and talker indices are omitted for the sake of readability.
    }
    \label{tab:dnn_configs}
\end{table}

\subsection{DNN architecture} 
We utilize the FT-JNF architecture proposed in~\cite{tesch_insights_2023} (see Fig.~\ref{fig:dnn_architecture}) with uni-directional LSTM layers. 
The architecture follows an STFT-based masking approach, of which we consider several variants with different microphone contributions to mask estimation and STFT masking. 
The details of each variant are listed in Table~\ref{tab:dnn_configs}.
The DNNs compute the complex-valued STFT masks $M_\mathrm{o}$ and/or 
$M_\mathrm{i}$, which are multiplied with the noisy STFTs $Y_\mathrm{i}$ and $Y_\mathrm{o}$ in a weighted overlap-add scheme.
If the DNN is not trained to output the mask $M_\mathrm{m}$ for microphone $m$, the corresponding channel is not used in filtering.
The DNN variants differ only in their input and output dimension. The first LSTM has 512, the second has 128 hidden units. The architecture variants consist of around 1.4M parameters each. 

\section{Results}
The results of the noise augmentation experiment are presented in Section~\ref{sec:noise_aug_results}. To investigate the contribution of each channel as auxilary or filtering input, the results of an ablation study are presented in Section~\ref{sec:channel_contribution_results}.  

\subsection{Noise augmentation}
\label{sec:noise_aug_results}

\begin{figure}[t]
    \centering
    \begin{subfigure}[b]{\columnwidth} 
    \includegraphics[width=\columnwidth]{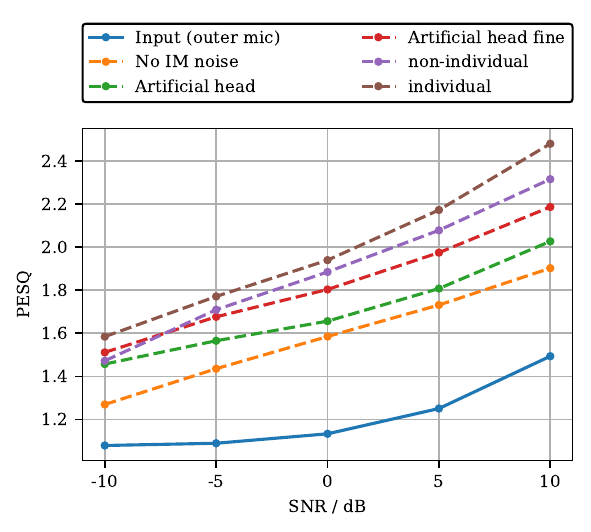}
        \label{fig:noise_aug_PESQ}
    \end{subfigure}
    \hfill
    \begin{subfigure}[b]{\columnwidth}
    \includegraphics[width=\columnwidth]{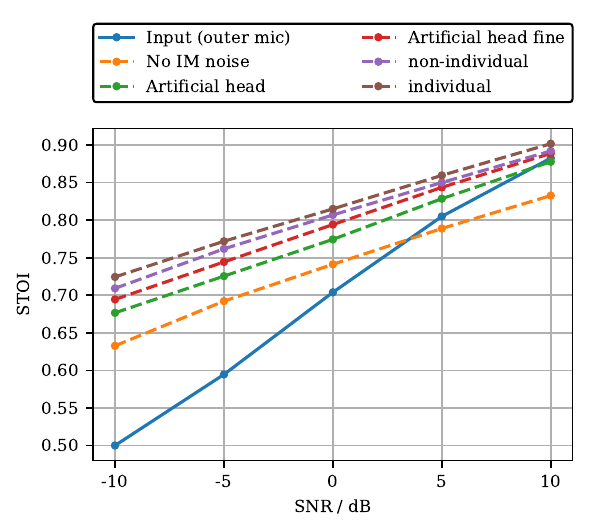}
        \label{fig:channel_use_STOI}
    \end{subfigure}
    \caption{Results of the noise augmentation experiment. 
    Here, the DNN variant which utilizes both outer and inner microphone in mask estimation and filtering (OM+IM) is used.} 
    \label{fig:results_noise_augmentation}
\end{figure}

The results of the noise augmentation experiment are shown in Fig.~\ref{fig:results_noise_augmentation}. 
If no IM noise is considered during training, the trained DNNs improve PESQ scores over the noisy OM signals. 
For low SNRs, the STOI of the processed signal is higher than of the noisy OM, but for high SNRs, it is lower.
Further improvement is gained by using AH RTFs for data augmentation instead of no incorporating IM noise during training.
When more directions are considered using fine instead of coarse resolution, the benefit is higher.
When non-individual talker RTFs are used instead of an AH, there is further improvement. 
Although the non-individual method utilizes less RTF measurements than the artificial head method with fine resolution, it achieves slightly higher scores. 
Finally, if individual RTFs are used in data augmentation, the highest own voice reconstruction performance is achieved. 
Overall, there is a consistent gain from simulating IM noise during training by using any of the proposed noise augmentation methods.

\subsection{Microphone contribution ablation study}
\label{sec:channel_contribution_results}

\begin{figure}[t] 
    \centering
    \includegraphics[width=\columnwidth]{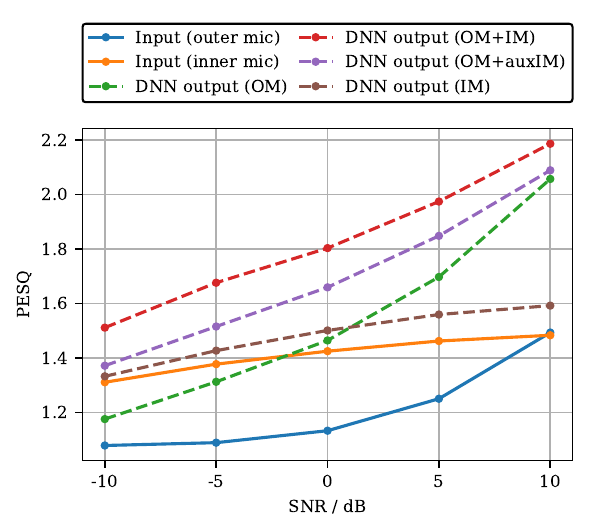}
    \caption{Channel contribution ablation study results. Augmented noise using AH RTFs with fine resolution was used for training.}
    \label{fig:results_channel_use}
\end{figure}

The results of the ablation study are shown in Fig.~\ref{fig:results_channel_use}.
The OM-DNN yields large improvements over the noisy OM signals in high SNRs, but smaller improvements in low SNRs. 
The performance of OM-DNN is better than IM-DNN for high SNRs, but worse for low SNRs. 
When both microphone signals are used as input for filtering, the OM+auxIM-DNN improves over the OM-DNN for all SNRs, while only yielding better scores than the IM-DNN above 0\,dB SNR. 
The OM+IM-DNN further improves performance over the OM+auxIM-DNN through the use of the IM signals as input in filtering.
Overall, we note a large benefit from using the IM in low SNRs, while in high SNRs the contribution of the OM is larger.

\section{Conclusion}
In this paper, we have proposed multi-microphone noise augmentation methods for DNN-based own voice reconstruction. 
Noise augmentation schemes for training a multi-microphone own voice reconstruction system were evaluated.
Experimental results show that incorporating noise augmentation in training of the considered own voice reconstruction system is beneficial.
Using individualized noise augmentation leads to the best performance. 
Additionally, we have investigated the SNR-dependent benefit of an IM, which is high especially in low SNRs.

\clearpage
\printbibliography

\end{document}